\documentclass{aa}
\usepackage{graphicx}
\usepackage{txfonts}
\bibpunct{(}{)}{;}{a}{}{,} 
\defcitealias{pal12}{PQM12}
\defcitealias{pal16}{Paper~I}
\defcitealias{men17}{Paper~II}

\begin{document}

\title{K-shell photoabsorption and photoionization of trace elements}

\subtitle{III. Isoelectronic sequences with electron number $19\leq N\leq 26$}

\author{C. Mendoza\inst{1}\thanks{Also Emeritus Research Fellow, IVIC, Caracas, Venezuela.}
  \and
  M.~A. Bautista\inst{1}
  \and
  P. Palmeri\inst{2}
  \and
  P. Quinet\inst{2,3}
  \and
  M.~C. Witthoeft\inst{4,5}
  \and
  T.~R. Kallman\inst{5}
}

\institute{Department of Physics, Western Michigan University, 1903 W Michigan Ave., Kalamazoo, MI 49008, USA \\
               \email{claudio.mendozaguardia@wmich.edu, manuel.bautista@wmich.edu}
   \and
   Physique Atomique et Astrophysique, Universit\'e de Mons - UMONS, 20 place du Parc, 7000, Mons, Belgium \\
              \email{patrick.palmeri@umons.ac.be, pascal.quinet@umons.ac.be}
   \and
   IPNAS, Universit\'e de Li\`ege, Campus du Sart Tilman, B\^at. B15, 4000, Li\`ege, Belgium
   \and
   ADNET Systems, Inc., Bethesda, MD 20817, USA
   \and
   NASA Goddard Space Flight Center, Greenbelt, MD, 20771, USA \\
              \email{michael.c.witthoeft@nasa.gov, timothy.r.kallman@nasa.gov}
}

\date{Received , ; accepted , }


\abstract
{This is the final report of a three-paper series on the K-shell photoabsorption and photoionization of trace elements (low cosmic abundance), namely F, Na, P, Cl, K, Sc, Ti, V, Cr, Mn, Co, Cu and Zn. K lines and edges from such elements are observed in the X-ray spectra of supernova remnants, galaxy clusters and accreting black holes and neutron stars, their diagnostic potential being limited by poor atomic data.}
{We are completing the previously reported radiative datasets with new photoabsorption and photoionization cross sections for isoelectronic sequences with electron number $19\leq N\leq 26$. We are also giving attention to the access, integrity and usability of the whole resulting atomic database.}
{Target representations are obtained with the atomic structure code {\sc autostructure}. Where possible, cross sections for ground-configuration states are computed with the Breit--Pauli $R$-matrix method ({\sc bprm}) in either intermediate or $LS$ coupling including damping (radiative and Auger) effects; otherwise and more generally, they are generated in the isolated-resonance approximation with {\sc autostructure}.}
{Cross sections were computed with {\sc bprm} only for the K ($N=19$) and Ca ($N=20$) isoelectronic sequences, the latter in $LS$ coupling. For the rest of the sequences ($21\leq N \leq 26$), {\sc autostructure} was run in $LS$-coupling mode taking into account damping effects. Comparisons between these two methods for K-like \ion{Zn}{xii} and Ca-like \ion{Zn}{xi} show that, to ensure reasonable accuracy, the $LS$ calculations must be performed taking into account the non-fine-structure relativistic corrections. The original data structures of the {\sc bprm} and {\sc autostructure} output files, namely photoabsorption and total and partial photoionization cross sections, are maintained but supplemented with files detailing the target ($N_T$-electron system, where $N_T=N-1$) representations and photon states ($N$-electron system).}
{We conclude that, due to large target size, the photoionization of ions with $N>20$ involving inner-shell excitations rapidly leads to untractable {\sc bprm} calculations, and is then more effectively treated in the isolated resonance approximation with {\sc autostructure}. This latter approximation by no means involves small calculations as Auger damping must be explicitly specified in the intricate decay routes.}


\titlerunning{K-shell photoabsorption of trace elements}
\authorrunning{C. Mendoza et al.}

\maketitle


\section{Introduction}

The present project is concerned with the computation of atomic data to improve the diagnostic capabilities of the spectral K lines and edges of chemical elements with low cosmic abundance (trace elements) -- namely F, Na, P, Cl, K, Sc, Ti, V, Cr, Mn, Co, Cu and Zn. In spite of their low abundances, they are nevertheless observed in the X-ray spectra of supernova remnants, galaxy clusters and accreting black holes and neutron stars \citep[see, for instance,][]{hwa00, mil06, bad08, kal09, tam09, ued09, nob10, par13}, from which they can be used to constrain the plasma characteristics. For this purpose level energies, radiative and Auger widths and fluorescence yields for K-vacancy levels in ions of the aforementioned isonuclear sequences were computed by \citet[hereafter \citetalias{pal12}]{pal12} with {\sc hfr} \citep[a Hartree--Fock code with relativistic corrections by][]{cow81}. Their atomic target representations were subsequently used by \citet[hereafter \citetalias{pal16}]{pal16} and \citet[hereafter \citetalias{men17}]{men17} to calculate with the Breit--Pauli $R$-matrix method \citep[{\sc bprm},][]{ber95} intermediate-coupling photoabsorption and photoionization cross sections for isoelectronic sequences with electron number $N\leq 18$. In this context, the resonance structures associated with the L and K ionization edges were studied in detail, particularly the radiation and spectator-Auger damping effects that lead to K-edge smearing \citep{pal02}.

In the present and final report of this project, we compute with the multi-purpose atomic structure code {\sc autostructure} $LS$-coupling photoabsorption and photoionization cross sections for trace ions in isoelectronic sequences with $19\leq N\leq 26$ assuming the isolated resonance approximation. This latter approach has been adopted because {\sc bprm} has proven to be ineffective due to the large target sizes involved. {\sc autostructure} accuracy has been previously shown to be reasonable for this purpose (see \citetalias{men17}), and is further examined here for the simpler K ($N=19$) and Ca ($N=20$) isoelectronic sequences.

An important part of this report is the curation aspects of the voluminous database that has been generated in this project, which is to become available online from the Centre de Donn\'ees astronomiques de Strasbourg (CDS\footnote{http://cdsweb.u-strasbg.fr/}). We emphasize, however, that the datasets have been computed with different numerical methods and angular couplings, and their content and structures may thus nominally vary.

\section{Numerical methods}
\label{methods}

The main task of the present project is to compute photoabsorption and total and partial photoionization cross sections for ions of the isonuclear series P, Cl, K, Sc, Ti, V, Cr, Mn, Co, Cu and Zn with electron numbers $N< Z-1$, where $Z$ is the atomic number identifying the sequence. As previously mentioned in \citetalias{pal16}, species with $Z-1\leq N\leq Z$ will be treated elsewhere. Cross sections for isoelectronic sequences with $N\leq 11$ were reported in \citetalias{pal16} and with $12\leq N\leq 18$ in \citetalias{men17}. We discuss here the details of the computations for fourth-row ions with $19\leq N\leq 26$.

The cross sections reported by \citetalias{pal16} and \citetalias{men17} were carried out in intermediate coupling (IC) with the relativistic (Breit--Pauli) {\sc bprm} method, which allows the inclusion of radiative and spectator-Auger damping \citep{rob95, gor96, gor00}. The target representations listed in Table~9 of \citetalias{pal12} were used but, for ions with $12\leq N\leq 18$, levels from the 2s-hole configurations $[{\rm 2s}]\mu$ were additionally included. Electronic orbitals were generated in a Thomas--Fermi--Dirac statistical potential with the {\sc autostructure} atomic structure code \citep{eis74, bad11}.

Due to the large target sizes required for species with $19\leq N\leq 26$, specially those bearing ground configurations ${\rm 3p^63d}^m$ with $m>2$, the {\sc bprm} approach is no longer practical. Exploratory calculations in \citetalias{men17} revealed that $LS$ cross sections could be adequately generated for such larger ions with a distorted-wave approach in the isolated-resonance approximation implemented in {\sc autostructure}. Target models are still those from \citetalias{pal12} but, as previously shown \citepalias{men17}, they are complemented with levels from configurations of the type $[{\rm 2s}]\mu$. Radiation damping is taken into account by {\sc autostructure} but, in contrast to {\sc bprm}, the Auger decay branches must be explicitly specified in the configuration list thus leading to lengthy calculations. The key feature of this approach is the decoupling of the photoionization and photoexcitation processes that enables the treatment of larger systems.

\section{Results}
\label{results}

\begin{figure}[!t]
  \centering
  \includegraphics[width=8.4cm]{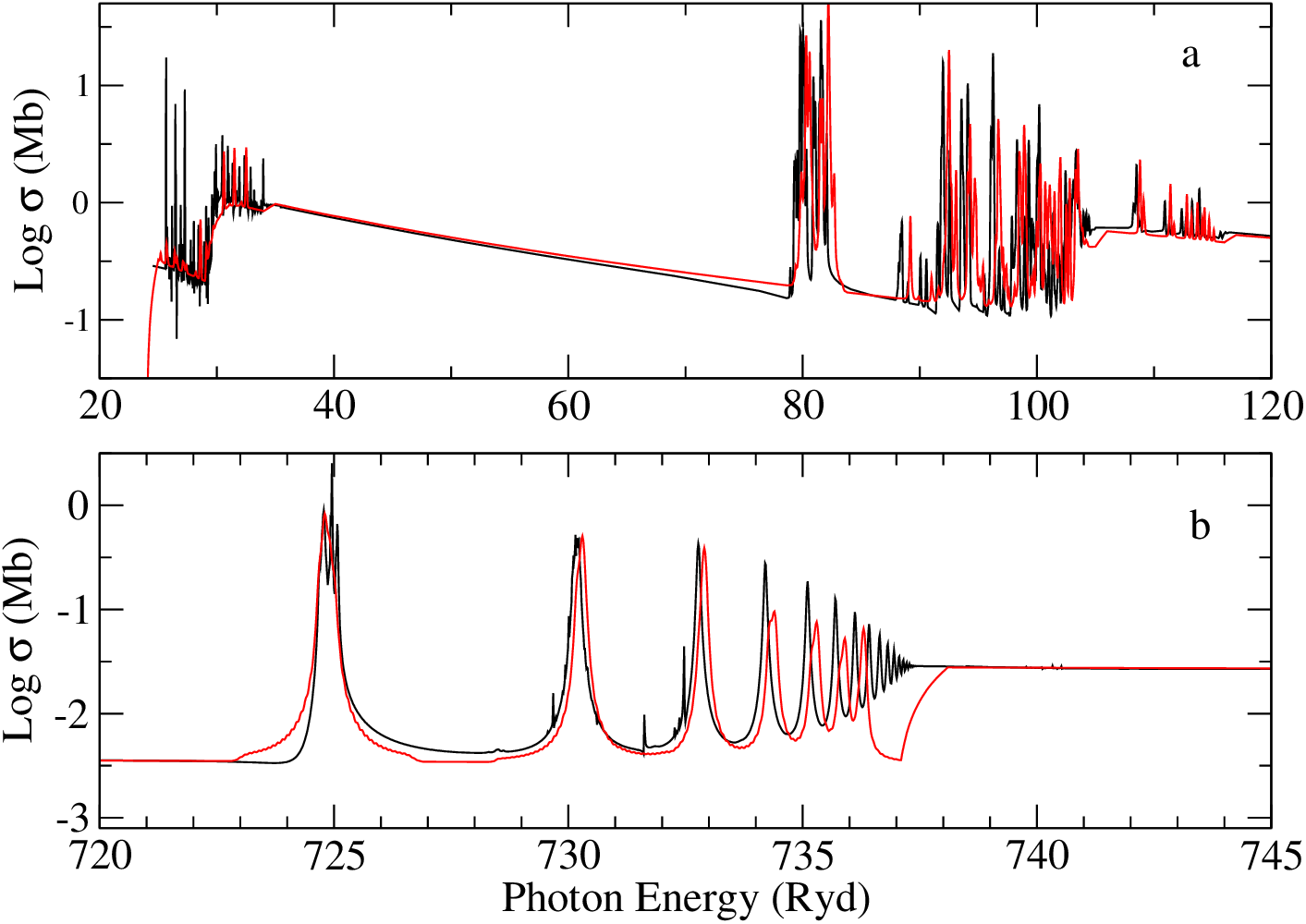}
  \caption{IC photoabsorption cross section of the ${\rm 3p^63d}\ ^2{\rm D}_{3/2}$ ground state of K-like \ion{Zn}{xii} computed with {\sc bprm} (black curve) and {\sc autostructure} (red curve). (a) Valence and L-edge regions; the {\sc bprm} cross section has been convolved with a Gaussian of width $\Delta E/E = 0.001$. (b) K-edge region; the {\sc autostructure} energy scale has been shifted by 0.9~Ryd to obtain resonance positional matching.} \label{znxii}
\end{figure}

For the trace elements there are hardly any previous multichannel photoabsorption cross sections for fourth-row ions with ground configurations ${\rm 3p^63d}^m$; therefore, cross sections are calculated from the ionization threshold up to the monotonic decreasing tail beyond the K edge. We illustrate the main findings of our analysis in terms of the Zn ions. For the potassium isoelectronic sequence ($N=19$) it is still possible, although involved, to perform a {\sc bprm} calculation in IC to compare with {\sc autostructure}. Such a comparison is carried out in Fig.~\ref{znxii} for the ${\rm 3p^63d}\ ^2{\rm D}_{3/2}$ ground state of \ion{Zn}{xii}, where the {\sc bprm} data below the L edge (${\sim}103$~Ryd) have been convolved for clarity with a Gaussian of width $\Delta E/E = 0.001$ (Fig.~\ref{znxii}a). Salient spectral features in this region are the M edge (${\sim}30$~Ryd) and the L$\alpha$ transition array at ${\sim}82$~Ryd, where resonance positions from the two numerical methods appear to be in adequate agreement. Fig.~\ref{znxii}b shows the near-K edge region above 720~Ryd, where the {\sc autostructure} energy scale has been shifted by 0.9~Ryd to ensure resonance positional matching. This discrepancy is a direct consequence of adjusting the {\sc bprm} thresholds manually to the energy values listed in Table~9 of \citetalias{pal12}, a procedure that was performed throughout for all ions with electron number $N\leq 18$. The broad dip at ${\sim}737$~Ryd in the {\sc autostructure} curve of Fig.~\ref{znxii}b is a numerical artifact resulting from the finite number ($n=10$) of K-vacancy states considered, and illustrates the difficulties of rendering properly the region just below threshold.

\begin{figure}[!h]
  \centering
  \includegraphics[width=8.4cm]{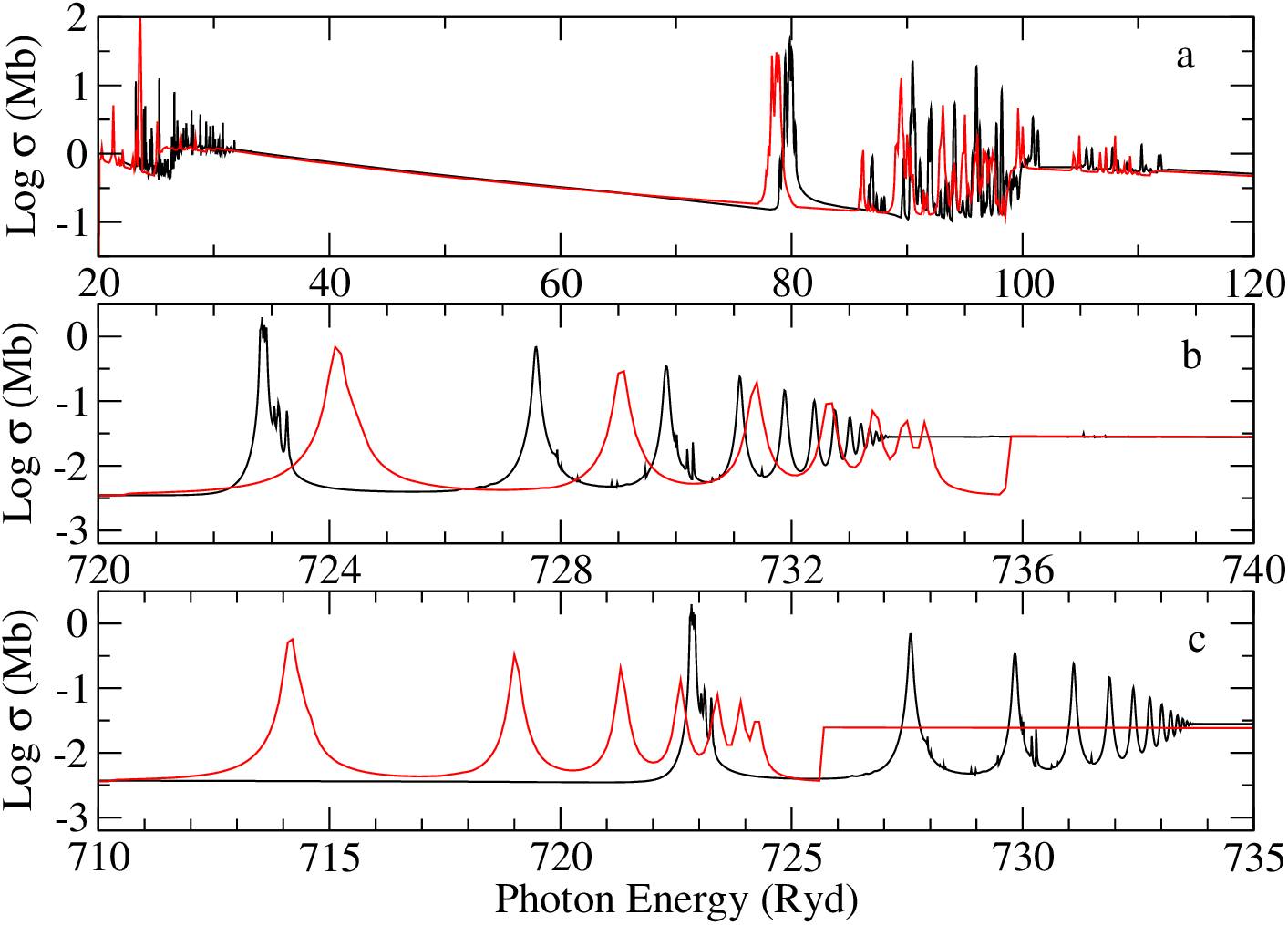}
  \caption{$LS$-coupling photoabsorption cross section of the ${\rm 3p^63d^2}\ ^3{\rm F}$ ground state of Ca-like \ion{Zn}{xi} computed with {\sc bprm} (black curve) and {\sc autostructure} (red curve). (a) Valence and L-edge regions; the {\sc bprm} cross section has been convolved with a Gaussian of width $\Delta E/E = 0.001$. (b) K-edge region. (c) K-edge region where the {\sc autostructure} cross section has been computed excluding the non-fine-structure relativistic corrections leading to a large positional discrepancy in the resonance structure.} \label{znxi}
\end{figure}

A similar situation is found in a comparison between {\sc bprm} and {\sc autostructure} in $LS$ coupling for the ${\rm 3p^63d^2}\ ^3{\rm F}$ ground term of Ca-like \ion{Zn}{xi} (Fig.~\ref{znxi}). The positions of the {\sc autostructure} L and M edges and the L$\alpha$ transition array are somewhat lower (${\sim}2$~Ryd) than {\sc bprm} (see Fig.~\ref{znxi}a) while the K lines are ${\sim}1.5$~Ryd higher (see Fig.~\ref{znxi}b). Such differences are found to be typical and manageable, which is not the case if the non-fine structure relativistic corrections are neglected in {\sc autostructure} as shown in Fig.~\ref{znxi}c, where now the discrepancies can be as large as 10~Ryd.

\begin{figure*}[!t]
  \centering
  \includegraphics[width=16cm]{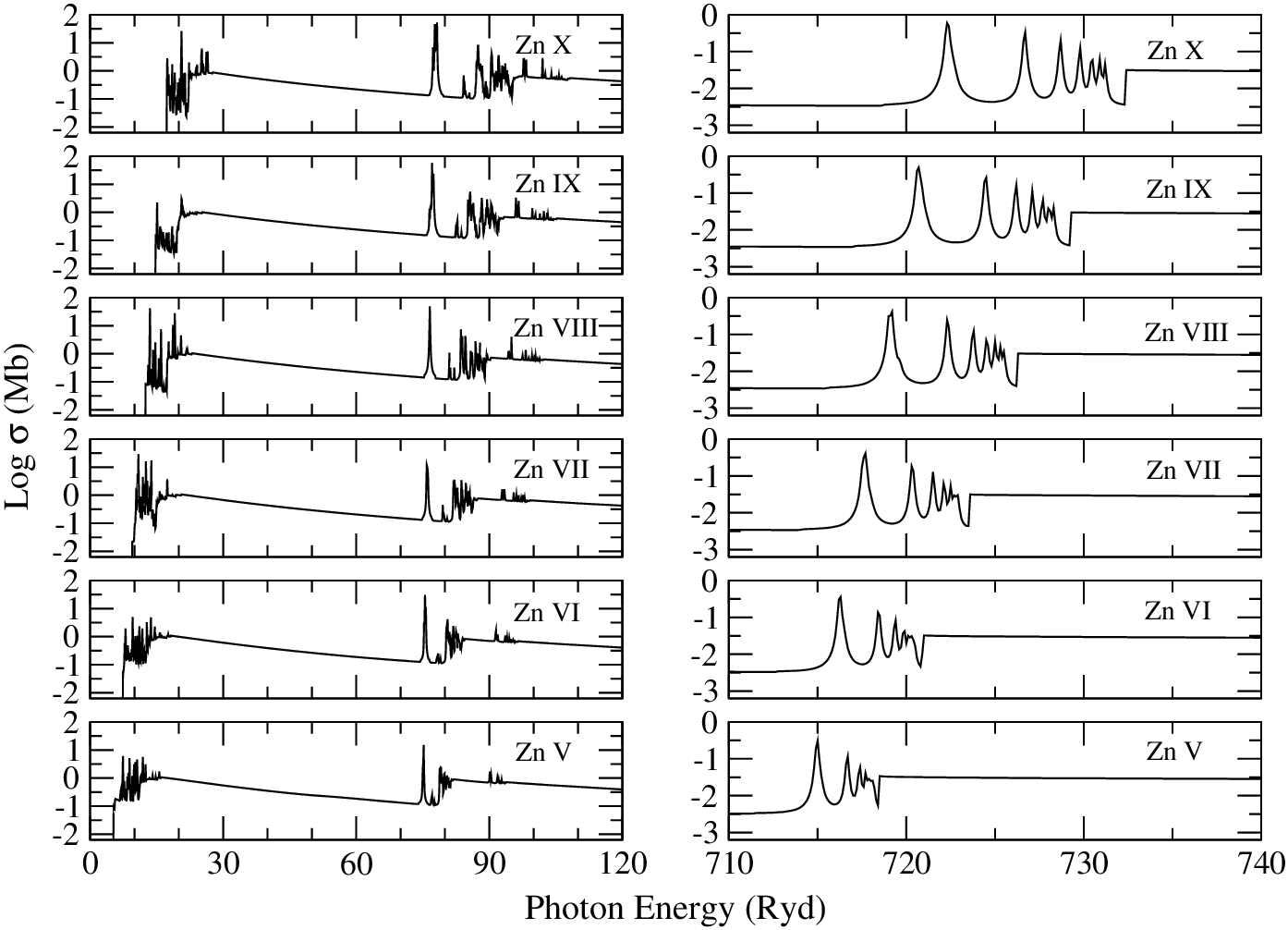}
  \caption{{\sc autostructure} $LS$ photoabsorption cross sections of the ground states of \ion{Zn}{x} (Sc-like),  \ion{Zn}{ix} (Ti-like), \ion{Zn}{viii} (V-like), \ion{Zn}{vii} (Cr-like), \ion{Zn}{vi} (Mn-like) and  \ion{Zn}{v} (Fe-like) in the valence and L-edge regions ({\it left column}) and K-edge region ({\it right column}). \label{mosaic}}
\end{figure*}

In Fig.~\ref{mosaic} we plot the photoabsorption cross sections of the Zn ions of the iron-group: \ion{Zn}{x} (Sc-like),  \ion{Zn}{ix} (Ti-like), \ion{Zn}{viii} (V-like), \ion{Zn}{vii} (Cr-like), \ion{Zn}{vi} (Mn-like) and  \ion{Zn}{v} (Fe-like). The left panels show the region below 120~Ryd dominated by the L and M edges and the L$\alpha$ transition arrays, while the right panels show the K edge and its associated damped lines, which are the spectral signatures of interest in X-ray astronomical spectra. An important issue here is how to improve the energy scale subject to the uncertainties in both the ionization potential and excitation thresholds. As shown by \citet{gat13b, gat13a} for the oxygen K lines, there are standing inconsistencies between the laboratory and observational energy scales that are deterrents in a reliable choice. As a consequence, no attempt is made here to adjust the energy scales.

\section{Database content and conventions}
\label{data}

As previously mentioned, the datasets generated in the present project are to become available online from the CDS. We give here a concise description of the adopted data conventions and structures to identify the files and facilitate the extraction of their content.

\subsection{Summary}

Following the close-coupling ($R$-matrix) convention, where a scattering system is conceived as a target plus a colliding electron, we identify the $N$-electron ionic species to be photoionized with the 2-tuple $(Z,N_T)$, where $Z$ is the atomic number and $N_T=N-1$ is the number of target electrons. In the present project we have been mainly concerned with photoabsorption and photoionization cross sections of states in the ionic ground configuration that will be referred to as {\em photon states}. The latter are identified by the 4-tuple $(0,2J,\pi,lev)$ in IC and $(2S{+}1,L,\pi,lev)$ in $LS$, where $J$ and $L$ are respectively the total and orbital angular momentum quantum numbers, $2S{+}1$ the spin multiplicity, $\pi$ the parity ($\pi=0$ for even, $\pi=1$ for odd) and $lev$ the level index within the $J\pi/SL\pi$ series. A summary of the database content in terms of these identifiers is given in Table~\ref{summary}. It may be seen therein that 24 isoelectronic sequences with target electron numbers $2\leq N_T\leq 25$ have been considered encompassing elements with atomic numbers $9\leq Z\leq 30$. Sequences with $N_T\geq 19$ have been treated in $LS$ coupling, and for $N_T\geq 20$, Ni species have also been included to complement previous work by \citet{wit11a}.

\subsection{Target data}

IC energy-level data for $(Z,N_T)$ targets in isoelectronic sequences with $2\leq N_T\leq 18$ and $LS$ energy-level data for those with $19\leq N_T\leq 25$ are listed in Table~\ref{target_IC} and Table~\ref{target_LS}, respectively. Energies in Rydbergs are given relative to the ground state, the latter listing the total ion energy. Table~\ref{target_IC} is very similar to Table~9 of \citetalias{pal12} but, since it is generated by {\sc bprm-stg3}, the level order may vary; moreover, for sequences with $N_T> 10$, the level number is greater since additional configurations were taken into account. Tables~\ref{target_IC}--\ref{target_LS} are essential when considering the partial photoionization cross sections that, for a specific $N$-electron photon state, are tabulated in $i$ order.

\subsection{Photon-state data}

The $N$-electron photon states for which cross sections have been computed are listed in Table~\ref{Nsystem_IC} (isoelectronic sequences with $3\leq N\leq 19$ in IC) and Table~\ref{Nsystem_LS} (sequences with $20\leq N\leq 26$, in $LS$); they essentially correspond to states within the ionic ground configuration. Energies are given in Rydbergs relative to the ionization potential. In systems with several photon states, IC cross sections for ions with $N\leq 19$ are tabulated in $i$ order.

\subsection{Photoabsorption and photoionization cross sections}

Cross sections have been computed using the serial version of the Breit--Pauli $R$-matrix codes for isoelectronic sequences with target electron numbers $N_T\leq 19$ and with {\sc autostructure} for $20\leq N_T\leq 25$ (see Table~\ref{summary}). Since the output files produced by these two suites of codes are somewhat different, we have kept the original nomenclature, structure and formats rather than making an attempt to unify them. We have also avoided the conversion of cross sections computed in $LS$ coupling to intermediate coupling. For an ionic species identified with the target tuple $(Z,N_T)\equiv ({\tt zz,nn})$, the following files have been included in the database.

For isoelectronic sequences with $2\leq N_T\leq 19$,
\begin{description}
  \item[{\tt zznn\_xpatot}\,:] total photoabsorption cross sections.
  \item[{\tt zznn\_xpisum}\,:] sum of the partial photoionization cross sections. It must be noted that this sum does not take into account the Auger damped component since it is treated in {\sc bprm} by means of a model potential that does not specify parental branching.
  \item[{\tt zznn\_xpipar}\,:] partial photoionization cross sections.
\end{description}

\noindent
For isoelectronic sequences with $20\leq N_T\leq 25$,
  \begin{description}
    \item[{\tt zznn\_xpitot}\,:] total photoabsorption cross sections.
    \item[{\tt zznn\_xdpisum}\,:] sum of the direct partial photoionization cross sections. It must be noted that in {\sc autostructure} the direct photoionization and photoexcitation processes are computed separately, and for the larger ions with $N_T> 19$, different target representations are implemented for each step that cannot be collated into a single unified target; therefore, the resonance component is neglected.
    \item[{\tt zznn\_xdpipar}\,:] direct partial photoionization cross sections.
  \end{description}

\section{Conclusions}
\label{conc}

Photoabsorption and photoionization cross sections have been computed for ions of the trace elements (elements with low cosmic abundance) F, Na, P, Cl, K, Sc, Ti, V, Cr, Mn, Co, Cu and Zn with electron number $N\leq 26$. Cross sections were computed in intermediate coupling  with the Breit--Pauli $R$-matrix method {\sc bprm} for sequences with $N\leq 20$;  for the rest of the sequences ($21\leq N \leq 26$), they were obtained, due to large target sizes, with {\sc autostructure} in $LS$ coupling assuming the isolated resonance approximation. Comparisons between these two methods for K-like and Ca-like ions show that, to ensure reasonable accuracy, it is vital to perform the $LS$ calculations taking into account the non-fine-structure relativistic corrections. Curation procedures performed on the datasets stored at the CDS have been described to facilitate their download and use. These datasets will also be processed to be included in the atomic database of the {\sc xstar} modeling code that calculates the physical conditions and emission spectra of photoionized gases \citep{bau01, kal09}.

As shown in Fig.~5b of \citetalias{men17}, the {\sc bprm} photoabsorption cross section of Ar-like \ion{Sc}{iv} in the K-edge region shows a small discontinuity at $\approx 331.5$~Ryd where the optical potential associated with Auger damping is switched on. This feature will be found in most {\sc bprm} curves in the K-edge region. Similarly, the discontinuous {\sc autostructure} cross section across the K-edge spectral head is due to the finite number of rendered resonances. This broad dip will be present in most of the cross sections computed with this code as appreciated in Fig.~\ref{znxii}b, Fig.~\ref{znxi}b,c and Fig.~\ref{mosaic} (right column) of the present report. Such numerical artifacts should be borne in mind.

There are hardly any previous calculations or experiments to reliably evaluate the accuracy of the present datasets. However, our target and collisional models have been extensively benchmarked with experiment and astronomical observations of K-line spectra involving ionic species from cosmic abundant isonuclear series and, in some cases, assessed by independent calculations with more refined approximations (e.g. the $R$-matrix plus pseudo-states framework). More precisely, we can cite the following sequences: N \citep{gar09, gha11, san11, sho13, gha14}; O \citep{gar05, gar11, gat13b, gat13a, gor13, mcl13b, mcl13a, gat14,mcl14, biz15, mcl17}; Ne, Mg, Si, S, Ar and Ca \citep{pal08a, wit11b}; Al \citep{pal11, wit13} and Ni \citep{pal08b, wit11a}.

The L-edge structure in ions with electron number $N > 12$, as discussed in \citetalias{men17}, is not expected to be fully converged in contrast to that of the K edge. Moreover, as shown by \citet{gor13} for \ion{O}{i}, theoretical K resonance positions are always subject to small wavelength adjustments to fit astronomical or laboratory measurements, which for this specific system have been shown to be discrepant. Therefore, the present data sets should be treated with due caution until they are independently verified.

\begin{acknowledgements}
This project is sponsored by NASA grant 12-APRA12-0070 through the Astrophysics Research and Analysis Program. PQ and PP are Research Director and Research Associate, respectively, of the Belgian Fund for Scientific Research F.R.S.-FNRS. We are indebted to Professor Nigel Badnell (Strathclyde University, UK) for innumerable communications regarding the possibilities, features and switches of the {\sc autostructure} computer package.
\end{acknowledgements}

\bibliographystyle{aa}



\begin{table*}[t]
  \caption{Summary of the ionic systems studied in the present work identified by the 2-tuple $(Z,N_T)$, where $Z$ is the atomic number and $N_T=N-1$ the number of target electrons. Photon states of the $N$-electron systems for which cross sections have been calculated are identified with the 4-tuple $(0,2J,\pi,elev)$ in IC and $(2S{+}1,L,\pi,elev)$ in $LS$. Isoelectronic sequences with $2\leq N_T\leq 18$ were treated in IC and those with $19\leq N_T\leq 25$ in $LS$.}
  \label{summary}
  \small
  \centering
  \begin{tabular}{r l l l}
  \hline\hline
  $N_T$ & $Z$ & $N$-electron photon states & Photon state identifiers \\
  \hline
  2  & $9,11,15,17,19,21{-}25,27,29,30$ & ${\rm 1s^22s\ ^2S_{1/2}}$ & $(0,1,0,1)$\\
  3  & $9,11,15,17,19,21{-}25,27,29,30$ & ${\rm 1s^22s^2\ ^1S_{0}}$ & $(0,0,0,1)$\\
  4  & $9,11,15,17,19,21{-}25,27,29,30$ & ${\rm 2s^22p\ ^2P^o_{1/2,3/2}}$ & $(0,1,1,1)$, $(0,3,1,1)$\\
  5  & $9,11,15,17,19,21{-}25,27,29,30$ & ${\rm 2s^22p^2\ ^3P_{0,1,2}, {^1D_2}, {^1S_0}}$ & $(0,0,0,1)$, $(0,2,0,1)$, $(0,4,0,1)$, $(0,4,0,2)$, $(0,0,0,2)$\\
  6  & $9,11,15,17,19,21{-}25,27,29,30$ & ${\rm 2s^22p^3\ ^4S^o_{3/2}, {^2D^o_{3/2,5/2}}, {^2P^o_{1/2,3/2}}}$ & $(0,3,1,1)$, $(0,3,1,2)$, $(0,5,1,1)$, $(0,1,1,1)$, $(0,3,1,3)$\\
  7  & $11,15,17,19,21{-}25,27,29,30$   & ${\rm 2s^22p^4\ ^3P_{0,1,2}, {^1D_2}, {^1S_0}}$ & $(0,0,0,1)$, $(0,2,0,1)$, $(0,4,0,1)$, $(0,4,0,2)$, $(0,0,0,2)$\\
  8  & $11,15,17,19,21{-}25,27,29,30$   & ${\rm 2s^22p^5\ ^2P^o_{1/2,3/2}}$ & $(0,1,1,1)$, $(0,3,1,1)$\\
  9  & $15,17,19,21{-}25,27,29,30$      & ${\rm 2s^22p^6\ ^1S_0}$ & $(0,0,0,1)$\\
  10 & $15,17,19,21{-}25,27,29,30$      & ${\rm 2p^63s\ ^2S_{1/2}}$ & $(0,1,0,1)$\\
  11 & $15,17,19,21{-}25,27,29,30$      & ${\rm 2p^63s^2\ ^1S_0}$ & $(0,0,0,1)$\\
  12 & $15,17,19,21{-}25,27,29,30$      & ${\rm 3s^23p\ ^2P^o_{1/2,3/2}}$ & $(0,1,1,1)$, $(0,3,1,1)$\\
  13 & $17,19,21{-}25,27,29,30$         & ${\rm 3s^23p^2\ ^3P_{0,1,2}, {^1D_2}, {^1S_0}}$ & $(0,0,0,1)$, $(0,2,0,1)$, $(0,4,0,1)$, $(0,4,0,2)$, $(0,0,0,2)$\\
  14 & $17,19,21{-}25,27,29,30$         & ${\rm 3s^23p^3\ ^4S^o_{3/2}, {^2D^o_{3/2,5/2}}, {^2P^o_{1/2,3/2}}}$ & $(0,3,1,1)$, $(0,3,1,2)$, $(0,5,1,1)$, $(0,1,1,1)$, $(0,3,1,3)$\\
  15 & $19,21{-}25,27,29,30$            & ${\rm 3s^23p^4\ ^3P_{0,1,2}, {^1D_2}, {^1S_0}}$ & $(0,0,0,1)$, $(0,2,0,1)$, $(0,4,0,1)$, $(0,4,0,2)$, $(0,0,0,2)$\\
  16 & $19,21{-}25,27,29,30$            & ${\rm 3s^23p^5\ ^2P^o_{1/2,3/2}}$ & $(0,1,1,1)$, $(0,3,1,1)$\\
  17 & $21{-}25,27,29,30$               & ${\rm 3s^23p^6\ ^1S_0}$ & $(0,0,0,1)$\\
  18 & $21{-}25,27,29,30$               & ${\rm 3p^63d\ ^2D_{3/2,5/2}}$ & $(0,3,0,1)$, $(0,5,0,1)$\\
     &                                  & ${\rm 3p^64s\ ^2S_{1/2}}$     & $(0,1,0,1)$\\
  19 & $22{-}25,27,29,30$               & ${\rm 3p^63d^2\ ^3F, {^3P}, {^1G}, {^1D}, {^1S}}$ & $(3,3,0,1)$, $(3,1,0,1)$, $(1,4,0,1)$, $(1,2,0,1)$, $(1,0,0,1)$\\
     &                                  & ${\rm 3p^63d4s\ ^3D, {^1D}}$ & $(3,2,0,1)$, $(1,2,0,2)$\\
  20 & $23{-}25,27{-}30$                & ${\rm 3p^63d^3\ ^4F, {^4P}, {^2H}, {^2G}}$ & $(4,3,0,1)$, $(4,1,0,1)$, $(2,5,0,1)$, $(2,4,0,1)$\\
     &                                  & ${\rm 3p^63d^3\ ^2F, {^2D2}, {^2D1}, {^2P}}$ & $(2,3,0,1)$, $(2,2,0,1)$, $(2,2,0,2)$, $(2,1,0,1)$\\
  21 & $24,25,27{-}30$                  & ${\rm 3p^63d^4\ ^5D, {^3H}, {^3G}, {^3F2}}$ & $(5,2,0,1)$, $(3,5,0,1)$, $(3,4,0,1)$,$(3,3,0,1)$\\
     &                                  & ${\rm 3p^63d^4\ ^3D, {^3F1}, {^3P2}, {^3P1}}$ & $(3,2,0,1)$, $(3,3,0,2)$, $(3,1,0,1)$,$(3,1,0,2)$\\
     &                                  & ${\rm 3p^63d^4\ ^1I, {^1G2}, {^1G1}, {^1F}}$ & $(1,6,0,1)$, $(1,4,0,1)$, $(1,4,0,2)$,$(1,3,0,1)$\\
     &                                  & ${\rm 3p^63d^4\ ^1D2, {^1D1}, {^1S2}, {^1S1}}$ & $(1,2,0,1)$, $(1,2,0,2)$, $(1,0,0,1)$,$(1,0,0,2)$\\
  22 & $25,27{-}30$                     & ${\rm 3p^63d^5\ ^6S, {^4G}, {^4F}, {^4D}}$ & $(6,0,0,1)$, $(4,4,0,1)$, $(4,3,0,1)$,$(4,2,0,1)$\\
     &                                  & ${\rm 3p^63d^5\ ^4P, {^2I}, {^2H}, {^2G2}}$ & $(4,1,0,1)$, $(2,6,0,1)$, $(2,5,0,1)$,$(2,4,0,1)$\\
     &                                  & ${\rm 3p^63d^5\ ^2G1, {^2F2}, {^2F1}, {^2D3}}$ & $(2,4,0,2)$, $(2,3,0,1)$, $(2,3,0,2)$,$(2,2,0,1)$\\
     &                                  & ${\rm 3p^63d^5\ ^2D2, {^2D1}, {^2P}, {^2S}}$ & $(2,2,0,2)$, $(2,2,0,3)$, $(2,1,0,1)$,$(2,0,0,1)$\\
  23 & $27{-}30$                        & ${\rm 3p^63d^6\ ^5D, {^3H}, {^3G}, {^3F2}}$ & $(5,2,0,1)$, $(3,5,0,1)$,$(3,4,0,1)$,$(3,3,0,1)$\\
     &                                  & ${\rm 3p^63d^6\ ^3F1, {^3D}, {^3P2}, {^3P1}}$ & $(3,3,0,2)$,$(3,2,0,1)$,$(3,1,0,1)$,$(3,1,0,2)$\\
     &                                  & ${\rm 3p^63d^6\ ^1I, {^1G2}, {^1G1}, {^1F}}$ & $(1,6,0,1)$,$(1,4,0,1)$,$(1,4,0,2)$,$(1,3,0,1)$\\
     &                                  & ${\rm 3p^63d^6\ ^1D2, {^1S2}, {^1D1}, {^1S1}}$ & $(1,2,0,1)$, $(1,0,0,1)$,$(1,2,0,2)$,$(1,0,0,2)$\\
  24 & $27{-}30$                        & ${\rm 3p^63d^7\ ^4F, {^4P}, {^2G}, {^2H}}$ & $(4,3,0,1)$, $(4,1,0,1)$, $(2,4,0,1)$, $(2,5,0,1)$\\
     &                                  & ${\rm 3p^63d^7\ ^2F, {^2D2}, {^2D1}, {^2P}}$ & $(2,3,0,1)$, $(2,2,0,1)$, $(2,2,0,2)$, $(2,1,0,1)$\\
  25 & $28{-}30$                        & ${\rm 3p^63d^8\ ^3F, {^3P}, {^1G}, {^1D}, {^1S}}$ & $(3,3,0,1)$, $(3,1,0,1)$, $(1,4,0,1)$, $(1,2,0,1)$, $(1,0,0,1)$\\
  \hline 
  \end{tabular}
\end{table*}

\begin{table*}
  \caption{IC energy-level data for target ions $(Z,N_T)$ with $N_T\leq 18$. The listed ground-state energy is the total ion energy while, for the rest of the levels, the level energy relative to the ground state is tabulated. Note: A complete version of this table is available electronically.}
  \label{target_IC}
  \small
  \centering                                                                                                                     \begin{tabular}{r r r c c c c l l r}                                                                                                       \hline\hline                                                                                                                   $Z$ & $N_T$ & $i$ & $2S+1$ & $L$ & $Pi$ & $2J$ & Conf & Term & $E$~(Ryd) \\
  \hline
  9	 & 2 & 1 & 1 & 0 & 0 & 0 & 1s2	 & 1s &	-151.119419 \\
  9	 & 2 & 2 & 3 & 0 & 0 & 2 & 1s.2s & 3s &	  53.127234 \\
  9	 & 2 & 3 & 3 & 1 & 1 & 0 & 1s.2p & 3p &	  53.770224 \\
  9	 & 2 & 4 & 3 & 1 & 1 & 2 & 1s.2p & 3p &   53.771590 \\
  9	 & 2 & 5 & 3 & 1 & 1 & 4 & 1s.2p & 3p &	  53.780248 \\
  9	 & 2 & 6 & 1 & 0 & 0 & 0 & 1s.2s & 1s &	  53.793005 \\
  9	 & 2 & 7 & 1 & 1 & 1 & 2 & 1s.2p & 1p &	  54.219478 \\
  11 & 2 & 1 & 1 & 0 & 0 & 0 & 1s2   & 1s &	-228.813549 \\
  11 & 2 & 2 & 3 & 0 & 0 & 2 & 1s.2s & 3s &	  81.425070 \\
  11 & 2 & 3 & 3 & 1 & 1 & 0 & 1s.2p & 3p &	  82.218026 \\
  11 & 2 & 4 & 3 & 1 & 1 & 2 & 1s.2p & 3p &	  82.222792 \\
  11 & 2 & 5 & 3 & 1 & 1 & 4 & 1s.2p & 3p &	  82.244728 \\
  11 & 2 & 6 & 1 & 0 & 0 & 0 & 1s.2s & 1s &	  82.269384 \\
  11 & 2 & 7 & 1 & 1 & 1 & 2 & 1s.2p & 1p &	  82.822648 \\
  \hline
  \end{tabular}
  \end{table*}

\begin{table*}
  \caption{$LS$ energy-level data for target ions $(Z,N_T)$ with $19\leq N_T\leq 25$. The listed ground-state energy is the total ion energy while, for the rest of the levels, the level energy relative to the ground state is tabulated. Note: A complete version of this table is available electronically.}
  \label{target_LS}
  \small
  \centering                                                                                                                     \begin{tabular}{r r r c c c l l r}                                                                                                       \hline\hline                                                                                                                   $Z$ & $N_T$ & $i$ & $2S+1$ & $L$ & $Pi$ & Conf & Term & $E$~(Ryd) \\
  \hline
   22 &  19 &   1 & 2 & 2 & 0 & 3p6.3d      & 2d  & -1701.498049 \\
   22 &  19 &   2 & 2 & 0 & 0 & 3p6.4s      & 2s  &     0.600370 \\
   22 &  19 &   3 & 4 & 2 & 1 & [3p]3d2     & 4d  &     2.268570 \\
   22 &  19 &   4 & 4 & 4 & 1 & [3p]3d2     & 4g  &     2.414790 \\
   22 &  19 &   5 & 4 & 1 & 1 & [3p]3d2     & 4p  &     2.425060 \\
   22 &  19 &   6 & 4 & 3 & 1 & [3p]3d2     & 4f  &     2.503180 \\
   22 &  19 &   7 & 2 & 2 & 1 & [3p]3d2     & 2d  &     2.523360 \\
   22 &  19 &   8 & 2 & 3 & 1 & [3p]3d2     & 2f  &     2.537130 \\
   22 &  19 &   9 & 2 & 1 & 1 & [3p]3d2     & 2p  &     2.591330 \\
   22 &  19 &  10 & 2 & 5 & 1 & [3p]3d2     & 2h  &     2.640820 \\
   22 &  19 &  11 & 2 & 4 & 1 & [3p]3d2     & 2g  &     2.660690 \\
   22 &  19 &  12 & 2 & 3 & 1 & [3p]3d2     & 2f  &     2.668140 \\
   22 &  19 &  13 & 4 & 2 & 1 & [3p]3d2     & 4d  &     2.725970 \\
   22 &  19 &  14 & 2 & 2 & 1 & [3p]3d2     & 2d  &     2.858760 \\
   22 &  19 &  15 & 4 & 0 & 1 & [3p]3d2     & 4s  &     2.889300 \\
   22 &  19 &  16 & 2 & 0 & 1 & [3p]3d2     & 2s  &     2.889300 \\
   22 &  19 &  17 & 2 & 4 & 1 & [3p]3d2     & 2g  &     2.904600 \\
   22 &  19 &  18 & 4 & 1 & 1 & [3p]3d.4s   & 4p  &     3.045990 \\
   22 &  19 &  19 & 2 & 1 & 1 & [3p]3d2     & 2p  &     3.059990 \\
   22 &  19 &  20 & 2 & 1 & 1 & [3p]3d.4s   & 2p  &     3.119540 \\
   22 &  19 &  21 & 4 & 3 & 1 & [3p]3d.4s   & 4f  &     3.156530 \\
  \hline
  \end{tabular}
  \end{table*}

\begin{table*}
  \caption{IC energy-level data for photon states of $(Z,N)$ systems for which cross sections have been computed ($N\leq 19$). Level energies are given relative to the ionization potential. Note: A complete version of this table is available electronically.}
  \label{Nsystem_IC}
  \small
  \centering                                                                                                                     \begin{tabular}{r r c c c c c l l r}                                                                                                       \hline\hline                                                                                                                   $Z$ & $N$ & $i$ & 0 & $2J$ & $Pi$ & $Lev$ & Conf & Term & $E$~(Ryd) \\
  \hline
   9 &	3 &	1 &	0 &	1 &	0 &	1 &	1s2.2s	& 2s &  -13.610382 \\
  11 &	3 &	1 & 0 & 1 & 0 & 1 & 1s2.2s	& 2s &  -22.040222 \\
  15 &	3 &	1 & 0 & 1 & 0 & 1 & 1s2.2s	& 2s &  -44.973266 \\
  17 &	3 &	1 & 0 & 1 & 0 & 1 & 1s2.2s	& 2s &  -59.496210 \\
  19 &	3 &	1 & 0 & 1 & 0 & 1 & 1s2.2s	& 2s &  -76.073846 \\
  21 &	3 &	1 & 0 & 1 & 0 & 1 & 1s2.2s	& 2s &  -94.756868 \\
  22 &	3 &	1 & 0 & 1 & 0 & 1 & 1s2.2s	& 2s & -104.878480 \\
  23 &	3 &	1 & 0 & 1 & 0 & 1 & 1s2.2s	& 2s & -115.529476 \\
  24 &	3 &	1 & 0 & 1 & 0 & 1 & 1s2.2s	& 2s & -126.713523 \\
  25 &	3 &	1 & 0 & 1 & 0 & 1 & 1s2.2s	& 2s & -138.434645 \\
  27 &	3 &	1 & 0 & 1 & 0 & 1 & 1s2.2s	& 2s & -163.519500 \\
  29 &	3 &	1 & 0 & 1 & 0 & 1 & 1s2.2s	& 2s & -190.819322 \\
  30 &	3 &	1 & 0 & 1 & 0 & 1 & 1s2.2s	& 2s & -205.243438 \\
  \hline
  \end{tabular}
  \end{table*}

\begin{table*}
  \caption{$LS$ energy-level data for photon states of the $(Z,N)$ systems for which cross sections have been computed ($20\leq N\leq 26$). Level energies are given relative to the ionization potential. Note: A complete version of this table is available electronically.}
  \label{Nsystem_LS}
  \small
  \centering                                                                                                                     \begin{tabular}{r r c c c c c l l r}                                                                                                       \hline\hline                                                                                                                   $Z$ & $N$ & $i$ & $2S+1$ & $L$ & $Pi$ & $Lev$ & Conf & Term & $E$~(Ryd) \\
  \hline
   22 &  20 & 1 & 3 & 1 & 0 & 1 & 3p6.3d2   & 3p & -2.612760 \\
   22 &  20 & 2 & 3 & 2 & 0 & 1 & 3p6.3d.4s & 3d & -2.358835 \\
   22 &  20 & 3 & 3 & 3 & 0 & 1 & 3p6.3d2   & 3f & -2.734018 \\
   22 &  20 & 4 & 1 & 0 & 0 & 1 & 3p6.3d2   & 1s & -2.299660 \\
   22 &  20 & 5 & 1 & 2 & 0 & 1 & 3p6.3d2   & 1d & -2.629597 \\
   22 &  20 & 6 & 1 & 2 & 0 & 2 & 3p6.3d.4s & 1d & -2.314143 \\
   22 &  20 & 7 & 1 & 4 & 0 & 1 & 3p6.3d2   & 1g & -2.585943 \\
   23 &  20 & 1 & 3 & 1 & 0 & 1 & 3p6.3d2   & 3p & -4.065680 \\
   23 &  20 & 2 & 3 & 2 & 0 & 1 & 3p6.3d.4s & 3d & -3.298038 \\
   23 &  20 & 3 & 3 & 3 & 0 & 1 & 3p6.3d2   & 3f & -4.210251 \\
   23 &  20 & 4 & 1 & 0 & 0 & 1 & 3p6.3d2   & 1s & -3.675538 \\
   23 &  20 & 5 & 1 & 2 & 0 & 1 & 3p6.3d2   & 1d & -4.083912 \\
   23 &  20 & 6 & 1 & 2 & 0 & 2 & 3p6.3d.4s & 1d & -3.253742 \\
   23 &  20 & 7 & 1 & 4 & 0 & 1 & 3p6.3d2   & 1g & -4.030104 \\
  \hline
  \end{tabular}
  \end{table*}

\end{document}